\documentclass[fleqn,twoside]{article}
\usepackage{espcrc2}

\usepackage{graphicx}

\newcommand{\la}{\lambda}

\newcommand{\mr}{\mathrm}
\newcommand{\mb}{\mathbf}

\def\fs{\; \; .}
\def\co{\; \; ,}

\newcommand{\til}{\tilde}

\newcommand{\cO}{\mathcal{O}}
\newcommand{\cF}{\mathcal{F}}

\newcommand{\ri}{\mr{i}}

\newcommand{\fpi}{F_{\pi}}

\newcommand{\epp}{\varepsilon_\pi}

\newcommand{\lp}{\lambda_\pi}

\newcommand{\AmS}{{\protect\the\textfont2
  A\kern-.1667em\lower.5ex\hbox{M}\kern-.125emS}}

\newcommand{\be}{\begin{equation}}
\newcommand{\ee}{\end{equation}}
\newcommand{\bea}{\begin{eqnarray}}
\newcommand{\eea}{\end{eqnarray}}

\title{The pion and proton mass in finite volume}

\author{Gilberto Colangelo\thanks{Speaker}, Andreas Fuhrer and Christoph
  Haefeli\\ 
Institut f\"ur Theoretische Physik der Universit\"at Bern\\
Sidlerstr. 5, 3012 Bern, Switzerland}
       
\begin{document}

\begin{abstract}
We calculate the finite volume effects for the pion and nucleon mass. For
the pion mass we present the results of a full two-loop calculation in
chiral perturbation theory. The outcome shows that the resummed version of
the L\"uscher formula we presented in an earlier work does indeed give an
excellent approximation to the two-loop result. In view of this result we
apply the same resummed formula to the nucleon mass. In the nucleon sector
the extension of the chiral expansion to higher quark masses appears to be
more problematic and it is therefore more difficult to make reliable
predictions for the size of the finite volume effects. We present some
preliminary numerical estimates.
\vspace{1pc}
\end{abstract}

\maketitle

\section{Introduction}
One of the many useful applications of effective field theories is to
estimate the size of systematic effects in lattice calculations. This
concerns both the unphysical size of the quark masses as well as the finite
lattice 
spacing which, in the framework of the effective field theory, can be seen
as a controlled modification of the underlying fundamental theory. Since
one can keep track of this modification even at the level of the effective
Lagrangian, one is able to calculate how the physics is changed at large
distances. A conceptually different problem is to estimate the effects due
to the finite volume -- in this case one is modifying the physics at large
distances but not touching the short-distance physics. Correspondingly, one
can evaluate the distortion due to the finite volume by calculating any
observable in infinite and in finite volume in the framework of the
effective Lagrangian \cite{Gasser:1987zq}. This evaluation can only be
reliable if the new scale introduced by the finite volume can be denoted as
a ``low energy scale'', i.e. if $L^{-1} \ll \Lambda$, where the latter is a
typical hadronic scale. In order to have a better numerical estimate of
how large a volume should be, we further observe that the minimum nonzero
momentum allowed in a box of size L is $p=2\pi/L$, and that if one
identifies $\Lambda$ with $4 \pi F_\pi$ (as is usually done in chiral
perturation theory) one obtains
\begin{equation}
\label{eq:2LF}
2 L F_\pi \gg 1 \qquad \Rightarrow \qquad L \gg 1 \mathrm{fm} \; \; .
\end{equation}
In recent years several analytical calculations of finite volume effects
within chiral perturbation theory (CHPT) have been performed by various groups
(for a recent review, see \cite{Colangelo:2004sc}). These results may be
used as a guideline by lattice groups when planning their runs and deciding
on the size of their volumes. Since these effects are generally small (at
the percent level) one could either try to estimate the volume for which
these effects are negligible (below one percent, say), or alternatively to
work in the smallest possible volume in which the effect can be reliably
calculated and correct analytically for the finite volume.
In the latter case it is of course important that the uncertainty in the
analytical calculation be carefully estimated. Moreover, before blindly
trusting some theoretical calculations, it is certainly a good practice to
check them at the claimed level of accuracy -- before one can save some
computer time on the volume at least some should be devoted to checking the
volume dependence.

In order to estimate the uncertainty in the finite volume calculations it
is necessary to check the convergence of the chiral expansion. This
requires evaluating at least two terms in the series for this specific
effect. Since finite volume effects start only at the one loop level, a
check of the convergence of the series requires a two-loop
calculation. Until recently \cite{Bijnens:2005ne,Haefeli:2005px} no such
calculations had ever been performed.  

A shortcut which did not require a full two-loop calculation but relied on
the L\"uscher formula for the masses \cite{Luscher:1985dn} (and on its
extension to decay constants \cite{Colangelo:2004xr}), was proposed in
\cite{Colangelo:2004sc,Colangelo:2005gd}. Asymptotic formulae {\em \`a
  la} L\"uscher express the finite volume shift as an integral over an
infinite volume amplitude -- {\em e.g.} the finite volume effect for the
pion (proton) mass is expressed as an integral over the $\pi \pi$ ($\pi N$)
scattering amplitude. Inserting the tree level representation for the
latter amplitude in the L\"uscher formula yields the asymptotically
dominating term of the one-loop finite volume calculation. But since the
relevant infinite-volume scattering amplitudes are usually known well
beyond the tree level, one can insert a better representation than the
leading order one and easily go beyond the leading order calculation of the
finite volume effect.

This has been done in \cite{Colangelo:2002hy,Colangelo:2003hf} for the pion
mass. The somewhat surprising results were: 1. the leading order in the
chiral expansion receives very large corrections at next-to-leading, but
reasonable ones at next-to-next-to leading order; 2. the leading
exponential term (as given by the L\"uscher formula) is numerically
dominating only where the finite volume effect becomes negligible. The
conclusions to be drawn from these results were that in order to reliably
calculate these effects it was necessary to go beyond the leading term both
in the chiral expansion as well as in the asymptotic expansion. The
possibility to use the NNLO representation for the $\pi \pi$ scattering
amplitude had however allowed us to show that, apart for the anomalously
large NLO correction (the reason for this large correction is well
understood, cf.~\cite{Colangelo:2003hf}), the series behaved as expected
and for not too large pion masses was clearly converging. The shift from
NLO to NNLO was typically smaller than the uncertainties in the NLO
calculation due to the low energy constants.

These results lead us to formulate a simple recipe \cite{Colangelo:2005gd}
to evaluate finite volume effects beyond LO. The recipe can be described as
a resummed L\"uscher formula: the integral over the infinite volume
amplitude which appears in the latter is generalized to one with an index
$n$ such that the asymptotic behaviour becomes $\exp(-\sqrt{n} M_\pi L)$,
and the full result is given by a sum over all $n$, with appropriate
coefficients.  This recipe yields the exact one-loop CHPT result if one
inserts the LO amplitude in the integral -- at NLO the correspondence is
not exact anymore, but we argued that we expected the resummed asymptotic
formula to give the largest part of the full two-loop calculation. Moreover
we could show that algebraically the formula improved the accuracy of the
asymptotic formula and that corrections were of the order
$O(e^{- \hat M L})$ with ($\hat M=(\sqrt{3}\!+\!1)/\sqrt{2} M_\pi$) instead
of the $O(e^{-\bar M L})$ with $\bar M=\sqrt{3/2} M_\pi$ of the plain
L\"uscher formula. 

Recently, we have completed the full two-loop calculation of the pion mass
and could confirm \cite{Haefeli:2005px} that the resummed asymptotic
formula is an excellent approximation to the full two-loop results, and can
therefore be used with confidence also in other cases. One interesting
application is that of the nucleon mass: the latter has been calculated to
one loop in CHPT \cite{AliKhan:2003cu} and the result has successfully
described lattice data, despite the fact that these had been obtained for
substantial pion masses and in rather small volumes. With the
resummed formula we can go beyond the one-loop calculation and better
estimate the uncertainties. These two new results (which will be published
soon) will be briefly described here.

\section{The pion mass to two loops}
In Ref.~\cite{Colangelo:2004sc,Colangelo:2005gd} we have proposed a
resummed version of the asymptotic formulae {\em \`a la} L\"uscher as an
efficient way to evaluate finite volume effects in CHPT beyond leading
order. The L\"uscher formula for the relative finite volume shift for the
pion mass reads as follows 
\bea
\label{eq:luscher}
R_{M_\pi}&\equiv& 
{M_\pi(L)-M_\pi \over M_\pi} = \\
&-&\frac{3}{16\pi^2 \la_\pi}
\int_{-\infty}^{\infty} \! d y \,\, \cF_\pi( \ri y) 
e^{-\sqrt{1 +  y^2}\,\la_\pi} \nonumber \\
&+& O(e^{-\bar M L}) \co\nonumber 
\eea
where $\cF_\pi(\nu)$ denotes the infinite volume forward ($t\!=\!0$)
$ \pi \pi$ scattering amplitude in Minkowski space, and 
\be
\lambda_\pi = M_\pi L 
\ee
is the box length in pion mass units. As discussed in the
introduction, $\lambda_\pi$ is assumed to be large, and the integral
in~(\ref{eq:luscher}) is only the dominating term in an asymptotic
expansion in exponentials of multiples of $\lambda_\pi$. The
estimated error in ~(\ref{eq:luscher}) is determined by $\bar M \geq
\sqrt{3/2} M_\pi$. In L\"uscher's proof of the formula the first step is to
show that the dominating contribution to finite volume effects comes from
diagrams where only one propagator is taken in finite volume. The propagator
in finite volume is
\be
G_L(x^0,\mb{x})=\sum_{\mb{n}}G_\infty(x^0,\mb{x}+\mb{n}L) \co
\label{eq:GL}
\ee 
where $G_\infty$ is the standard, infinite-volume propagator. The term
with $\mb{n}=\mb{0}$ in the sum is simply the infinite-volume propagator.
According to the rules for doing CHPT calculations in finite volume
\cite{Gasser:1987zq}, one has to proceed as in infinite volume and simply
write all propagators as in~(\ref{eq:GL}). As shown by L\"uscher the
dominating term comes from graphs where one propagator at a time is taken
in finite volume. He then showed that in this class of diagrams the
dominating contribution comes from the terms with $|\mb{n}|=1$ in the sum
in~(\ref{eq:GL}). Contributions from all other terms in the sum
in~(\ref{eq:GL}), however, will also show up in the complete CHPT
calculation and can be dealt with exactly as the first term in the sum.
This is the origin of the resummation we suggested
in~\cite{Colangelo:2004sc,Colangelo:2005gd} and leads to the resummed
formula 
\bea 
\label{eq:resum}
R_{M_\pi}&=&-{1 \over32\pi^2\lambda_\pi} \!
\sum_{n=1}^{\infty}\!
{m(n)\over\sqrt{n}} \times  \\
&& \hspace*{-4mm}\int\limits_{-\infty}^\infty\!dy\; \cF_\pi(\ri
y)\,e^{-\sqrt{n(1+y^2)}\lambda_\pi} + O(e^{-\bar M L}) 
\nonumber
\eea 
where $m(n)$ is the number of integer vectors $\bf{z}$ with ${\bf{z}}^2=n$.
It is easy to verify that the term with $n=1$ is identical to the original
L\"uscher formula~(\ref{eq:luscher}). Notice that although the formula
contains subleading terms in the asymptotic expansion, the algebraic
accuracy of the formula is in principle the same. On the other hand, an
analysis of the contributions at the two-loop level not included in the
resummed asymptotic formula shows that at this order in the chiral
expansion \cite{Colangelo:2005gd} the formula receives corrections only of
order $\exp(- \hat M L)$ with $\hat M = M_\pi(\sqrt{3}\!+\!1)/\sqrt{2}$,
and so is also algebraically improved with respect to the original
L\"uscher formula. 

On the basis of this and other arguments we have used this resummed formula
to make a thorough numerical analysis of finite volume effects for masses
and decay constants of the pseudoscalar mesons \cite{Colangelo:2005gd}.
It is important, however, to check the claim that the resummed formula
provides the main contribution of a full NLO calculation of the finite
volume effects, at least in one concrete example. We have now done
that by calculating the pion mass to two loops in CHPT.

The pion mass in finite volume to two loops can be written as
\[
  M_{\pi}(L)^2 = M_\pi^2 -\Sigma^{(1)}- \Sigma^{(2)} , \; \; M_\pi^2 =
  M^2 - \Sigma^{(0)} \; ,
\]
where
\begin{eqnarray}
  \label{eq:sum2}
  \Sigma^{(1)} &=& I_p + I_c + \cO(\xi^3) \co \\[1.5ex]
    I_p        &=& \!\!\!
    {M_\pi^2\over16\pi^2\lambda_\pi} \! \sum_{n=1}^{\infty}\!
    {m(n)\over\sqrt{n}} \times \nonumber \\
  && \int\limits_{-\infty}^\infty\!dy\;
  \cF_\pi(\ri y)\,e^{-\sqrt{n(1+y^2)}\lambda_\pi} \co \nonumber \\[1.5ex]
    I_c        &=&
  -\frac{\ri M_\pi^2}{32\pi^3\lambda_\pi} 
  \sum_{n=1}^{\infty} \frac{m(n)}{\sqrt{n}} \times \nonumber \\
  &&  \!\!\! \!\!\!  \!\!\! 
  \int\limits_{-\infty}^{\infty} \!\!\! dy
  \int\limits_{4}^{\infty} \!\!\! d\til s\,
  \frac{e^{-\sqrt{n(\til s+y^2)}\lambda_\pi}}
  {\til s+2\ri y}\,
  \mr{disc}\big[\cF_\pi(\til s,\!1\!+\!\ri y\!)\big] \co \nonumber
\end{eqnarray}
and introduced the abbreviation
\be
  \label{3eq:xi}
  \xi = \frac{M_\pi^2}{16\pi^2 \fpi^2}
  \fs
\ee
The expression for $\Sigma^{(1)}$ shows that at the two-loop level there
are contributions from diagrams with only one propagator in finite volume
which are not included in the resummed formula (\ref{eq:resum}). We have
denoted these as $I_c$ and shown that these can also be represented
compactly, as a double integral over the off-shell $\pi \pi$ scattering
amplitude. 

Contributions from diagrams where more than one propagator is taken in
finite volume are by definition not included in the resummed L\"uscher
formula. The whole $\Sigma^{(2)}$ contribution is a two-loop effect which
goes beyond the latter and can be represented as
\begin{eqnarray}
  \Sigma^{(2)} &=& \frac{M_\pi^2 \xi^2}{8} \left[ 
   9\til{g}_1(\lambda_\pi)^2-\lambda_\pi\, \til{g}_1(\lambda_\pi)
   \til{g}_1'(\lambda_\pi)\right] \nonumber \\
           && +M_\pi^2 \xi^2 \Delta +\cO(\xi^3) \co
\end{eqnarray}
where $\til{g}_1$ is the one-loop tadpole graph. The part from the sunset type
diagrams which can not be written in terms of $\til{g}_1$ is denoted here with
$\Delta$. It can be written as double integrals and can be evaluated
numerically.  

\begin{figure}[t]
\includegraphics[width=0.48\textwidth]{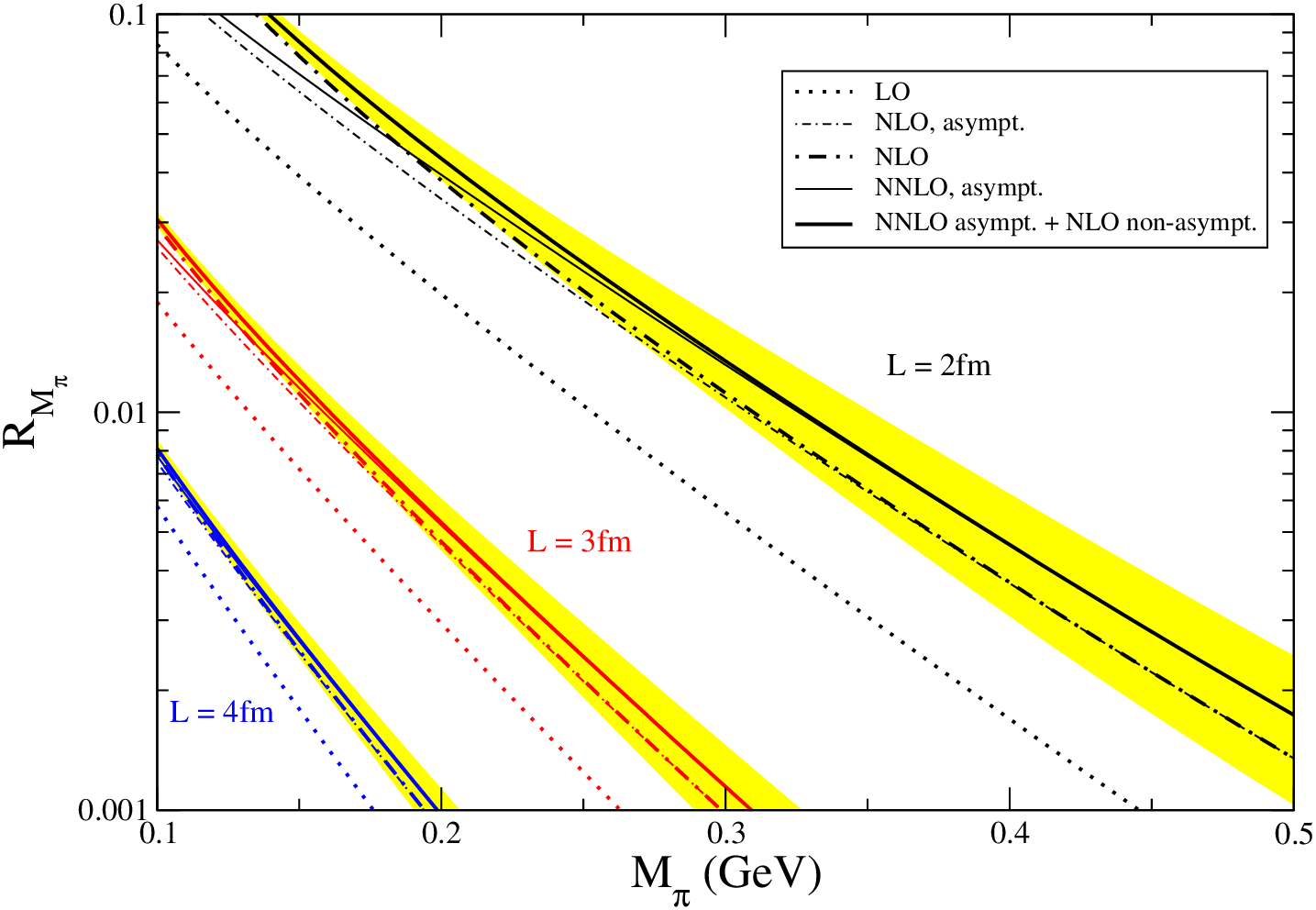}
\caption{\label{fig:2lmpi}Relative finite volume effect for the pion mass
  as a function of the pion mass for $L=2,3$ and $4$ fm.}
\end{figure}

The numerical analysis of this formula is shown in Fig.~\ref{fig:2lmpi}. It
is evident that the contributions which go beyond the resummed L\"uscher
formula are sizeable only in the region where $M_\pi L$ is not large, {\em
  i.e.} where the framework in which we are doing our calculations is not
reliable anymore.  The results show that the resummed L\"uscher formula
provides an efficient and reliable way to go beyond leading order in the
evaluation of finite volume effects.

\section{The nucleon mass with the resummed L\"uscher formula}
In this section we apply the resummed L\"uscher formula to the
nucleon mass and first provide the relevant analytical expression. The
difference to the case of the pion mass is that now the $\pi N$ scattering
amplitude, instead of the $\pi \pi$, has to be inserted in the
integral. 

The formula for the relative finite volume correction
$R_N=(m_N(L)-m_N)/m_N$ reads as follows 
\begin{eqnarray}
R_N\!\!&\!\!=\!\!&\!\! \frac{3 \epp^2}{4 \pi^2 } \sum_{n=1}^{\infty}
\frac{m(n)}{\sqrt{n} \lp}\bigg[ 2 \pi \epp g_{\pi N}^2
e^{-\sqrt{n(1-\epp^2)}\lp} \nonumber \\
&-& \!  \int_{-\infty}^{\infty} \! dy
e^{-\sqrt{n(1+y^2)}\lp} \tilde D^{+}(y) \bigg] \co
\label{eq:RN}
\end{eqnarray}
where
\begin{equation}
\epp=\frac{M_\pi}{2 m_N}\; , \; \; \tilde D^{+}(y)=
m_N D^+(i M_\pi y,0) \co
\label{eq:defs}
\end{equation}
where the latter is one of the components of the elastic $\pi N$ scattering
amplitude, which is defined as follows:
\begin{equation}
\begin{array}{l}
T(\pi^a(q) N(p) \to \pi^{a'}(q') N(p'))\equiv T_{a'a}= \\
\qquad \qquad \qquad \qquad =\delta_{a'a}T^++\frac{1}{2} [\tau_{a'},\tau_a]
T^- \fs 
\end{array}
\end{equation}
Each of the two isospin components, $T^+$ and $T^-$ is then broken down
into 
\begin{equation}
T^\pm=\bar u'\left[ D^\pm(\nu,t)-\frac{1}{4 m_N} [ q \hskip -0.2cm/',q
  \hskip -0.2cm / ] B^\pm(\nu,t) \right] u
\end{equation}
and each of the amplitudes depends on two kinematical variables:
\begin{equation}
\begin{array}{l}
t=(q-q')^2 \; , \; \;  \nu=\frac{s-u}{4 m_N}\co\\
s=(p+q)^2\; , \; \; u=(p-q')^2 \fs
\end{array}
\end{equation}

As seen in (\ref{eq:RN},\ref{eq:defs}), the $\pi N$ amplitude is needed
here in a particular kinematical configuration, namely for $t=0$ and for
$\nu$ purely imaginary and small: the contributions with large values of
$\nu$ are suppressed by the exponential weight in the integral in
(\ref{eq:RN}). It is therefore natural to make a Taylor expansion of
the amplitude around $\nu=0$ after having subtracted the pole due to the
one-nucleon exchange diagram (also called the Born term). Such an expansion
is in fact already well known in the phenomenology and is referred to as
the subthreshold expansion. It reads as follows
\be
D^+(\nu,0)=D^+_\mathrm{pv}(\nu,0)+D^+_\mathrm{p}(\nu,0)+D^+_\mathrm{na}(s,u)
\co
\ee
where
\bea
D^+_\mathrm{pv}(\nu,0)&=&\frac{g_{\pi N}^2}{m_N}
\frac{\nu_B^2}{\nu_B^2-\nu^2} \nonumber \co\\
D^+_\mathrm{p}(\nu,0)&=&d^+_{00}+d^+_{10}\nu^2+d^+_{20} \nu^4\co
\eea
and $\nu_B=-M^2_\pi/(2m_N)$. The function $D^+_\mathrm{na}(s,u)$ contains the
analytically nontrivial part of the amplitude. Up to order
$q^4$ in the chiral expansion this can be written as a sum of two single
variable functions: 
\be
D^+_\mathrm{na}(s,u)=D^+_1(s)+D^+_1(u)+O(q^5)
\ee
which admit the following dispersive representation:
\be
D^+_1(s)=\frac{\nu^5}{\pi} \int_{M_\pi}^\infty d \nu' \frac{
  \mathrm{Im} D_1^+(s')}{\nu^{\prime 5} (\nu'-\nu -i \epsilon)}
\co
\ee
where $s'=2m_N \nu'+m_N^2+M_\pi^2$. This representation shows that, due to
the large number of subtractions, the function $D_1^+$ is small near
$\nu=0$. Its contribution to the finite size shift of
the nucleon mass is negligible. 

This observation leads us to the following expression for the relative
finite volume shift $R_N$:
\bea
R_N&=&\frac{3 \epp^2}{4 \pi^2} \sum_{n=1}^\infty \frac{m(n)}{\sqrt{n} \lp}
\Bigg[ 2 \pi g_{\pi N}^2 \epp  e^{-\sqrt{n(1-\epp^2)}\lp} \nonumber \\
&-& g_{\pi N}^2 \epp^2 I_\mathrm{pv}(\sqrt{n} \lp)  - \bar d^+_{00}
B^0(\sqrt{n} \lp) \nonumber  \\
&+& \bar d^+_{10} B^2(\sqrt{n} \lp) - \bar
d^+_{20} B^4 (\sqrt{n} \lp) \Bigg]  \nonumber \\
&+& R_{N,\mathrm{na}} 
\co
\label{eq:RNb}
\eea
where
\bea
I_\mathrm{pv}(\lp)&=&\int_{-\infty}^\infty dy
\frac{e^{-\sqrt{(1+y^2)}\lp}}{\epp^2+y^2}  \; , \\
B^k(\lp)&=&\int_{-\infty}^\infty dy \; y^k e^{-\sqrt{(1+y^2)}\lp}
\nonumber 
\eea
are the relevant finite volume integrals, $\bar d^+_{i0}=m_N M_\pi^{2i}
d^+_{i0}$ and $R_{N,\mathrm{na}}$ is the remainder coming from the
(subtracted) analytically nontrivial part of the amplitude.

If we neglect the contribution $R_{N,\mathrm{na}}$ the
representation~(\ref{eq:RNb}) is very simple and expresses the finite
volume shift of the nucleon mass in terms of only a handful of physical
observables: the pion and proton masses, $M_\pi$, $m_N$, the pion-nucleon
coupling constant $g_{\pi N}$ and the three subthreshold parameters $\bar
d^+_{i0}$ (the latter are not directly observable, but can be obtained from
data with some theoretical treatment, cf.~\cite{Hohler:1984ux}). If one
knows the low-energy constants (LEC) which appear in the chiral
representation of these quantities, one can predict their quark mass
dependence and therefore the finite volume shift $R_N$ as a function of the
quark masses.

An explicit representation for the quark mass dependence of the quantities
which appear in Eq.~(\ref{eq:RNb}) up to $O(p^4)$ can be found in
\cite{Becher:2001hv}. Unfortunately, however, our knowledge of the LEC
which appear in there is much worse than that for the LEC of $\pi \pi$
scattering. In short we can say that the quark mass dependence pf the $\pi
N$ scattering amplitude, even in the subthreshold region, {\em i.e.} far
away from the physical singularities of the scattering amplitude, is not
very well known. We stress that for the physical value of the quark masses
we do have reliable phenomenological information about the quantities which
appear in Eq.~(\ref{eq:RNb}), and that the finite volume effects can be
therefore evaluated with rather small uncertainties,
cf.~\cite{Koma:2004wz}. The problem is the extrapolation to higher quark 
masses.

We have investigated this question and adopted the following strategy: we
fit the physical values of the five quantities $m_N$, $g_A$ and the three
subthreshold parameters $d_{i0}$ (cf.~\cite{Hohler:1984ux}) as well as the
lattice calculations of $m_N$ (cf.~\cite{Aoki:2002uc,AliKhan:2003cu}) and
$g_A$~\cite{galatt} for $M_\pi \sim 0.5$ GeV. This gives us seven
constraints and allows us to fix seven of the LEC appearing in the formulae
for the mass dependence. On some of the other LEC there is phenomenological
information coming from $\pi N$ or $\pi N \to \pi\ \pi N$ scattering --
whenever possible we use this also. Some of the LEC, however, remain
unconstrained and have to be estimated on a purely theoretical basis.  The
details of this analysis, including a discussion of the numerical values of
the various LEC, will be discussed elsewhere \cite{CF}.

An important point we wish to emphasize here is that our analysis indicates
that the Born term is dominating even at higher values of pion masses. This
means that the most important information we need, concerns the quark mass
dependence of $m_N$ and $g_{\pi N}$. For the former we rely on available
lattice calculations (although at somewhat high pion masses), whereas for
the latter we rely on the Goldberger--Treiman relation:
\be
g_{\pi N}=\frac{g_A m_N}{F_\pi} \left(1- \frac{2 d_{18} M^2}{g}+
  \mathcal{O}(M^4) \right)
\label{eq:GT}
\ee 
and fix the LEC $d_{18}$ from the phenomenology. This gives a mild
quark mass dependence in the relation -- much more important is the quark
mass dependence of $g_A$ and $m_N$. Also for $g_A$ we rely on available
lattice calculations (again at somewhat high pion masses), which indicate a
mild dependence of the axial charge on the quark mass, and try to
interpolate the physical and the lattice value with the chiral
representation. In this case, however, the phenomenological information
about the LEC (which determine the quark mass dependence of $g_A$) would
actually allow us to predict the latter \cite{Becher:2001hv}:
\bea
g_A \!\!&\!\!=\!\!&\!\! g +\left(4 \tilde d_{16}-\frac{g^3}{16\pi^2 F^2}
  \right) M^2 + \frac{(1+g^2)g M^3}{8 \pi mF^2}  \nonumber \\
&\!\!-\!\!&\!\! \frac{(c_3-2 c_4)g M^3 }{6 \pi F^2} 
 + \mathcal{O}(M^4) \fs
\eea
The constant $\tilde d_{16}$ has been determined from $\pi N \to \pi \pi N$
scattering measurements \cite{Fettes:1999wp} to be between $-1$ and $-2$
GeV$^{-2}$, with about $1$ GeV$^{-2}$ uncertainty\footnote{The published
  values of $d_{16}$ have later been corrected in the final version of the
  PhD thesis of Nadia Fettes (see also the discussion in
  \cite{Hemmert:2003cb}). The numerical change, however, does not change
  the qualitative picture.}, whereas the combination $c_3 -2 c_4$ can be
determined from the subthreshold parameters of $\pi N$ scattering
\cite{Becher:2001hv} and comes out to be also negative, with a value around
$-9$ GeV$^{-2}$. Unfortunately the prediction indicates a very strong quark
mass dependence already at very small quark masses and makes it difficult
to extrapolate to the region where the lattice data are obtained: the
$O(M^2)$ correction is large and negative and the $O(M^3)$ large and
positive. Of course one can choose the value of $g_A$ in the chiral limit
such that the value for the physical pion mass comes out right, but as soon
as one goes to higher pion masses, $g_A$ tends to explode. The lattice
data, on the other hand, indicate that at $M_\pi \sim 0.5$ GeV, $g_A$ is
somewhat lower, but not by much, than the physical value. We have to
conclude that either the phenomenological determination of the LEC
appearing in $g_A$ is unreliable, or that the region of quark masses where
the chiral expansion works for $g_A$ is very small.

The determination of $d_{16}$ is indeed subject to large uncertainties (of
the order of 50 to 80\% according to the estimates of Fettes, which do not
seem to include any systematic effects), but those of $c_3$ and $c_4$ are
more solid and indicate a small region of convergence of the chiral
expansion for $g_A$. The presence of these large corrections has indeed
been known since a long time (cf. \cite{Kambor:1998pi}) and a possible
solution of the discrepancy with the observed mass dependence (or the lack
thereof) in lattice calculations has already been proposed by Hemmert,
Procura and Weise \cite{Hemmert:2003cb}. In this paper the $\Delta$
resonance is included explicitly in the calculation and it is shown that if
one stops at order $\epsilon^3$ in the small-scale expansion (SSE) one
obtains a relatively steep dependence only close to the chiral limit,
whereas at higher quark masses $g_A$ is a flat function of $M_\pi$.  While
this analysis identifies a mechanism that provides a change of behaviour
between the very low and the middle pion mass region, it does not provide a
fully satisfactory understanding of the mass dependence of $g_A$, in our
opinion. The constant $d_{16}$ occurs also in the SSE as an independent
LEC, in fact unrelated to the $\Delta$ resonance. In \cite{Hemmert:2003cb}
the value of $d_{16}$ is also taken from the work of Fettes et al.
\cite{Fettes:1999wp}: this means that even in the SSE the term of order
$M^2$ is a large correction already at the physical pion mass, and that the
size of this correction grows fast with the pion mass.  The contributions
from the $\Delta$, which are also large, compensate these large corrections
and give a rather flat behaviour up to $0.5$ GeV.  However, since the
$\Delta$ and the $d_{16}$ contributions are physically unrelated (also in
the framework of the SSE), this compensation appears to be the result of a
fine tuning between the value of $d_{16}$ and the $\Delta$ resonance
couplings. The analysis in \cite{Hemmert:2003cb} shows that the $\Delta$
contributions could tame the strong dependence of $g_A$ on the quark mass,
but does not yet explain why this happens.

\begin{figure}[t]
\includegraphics[width=0.48\textwidth]{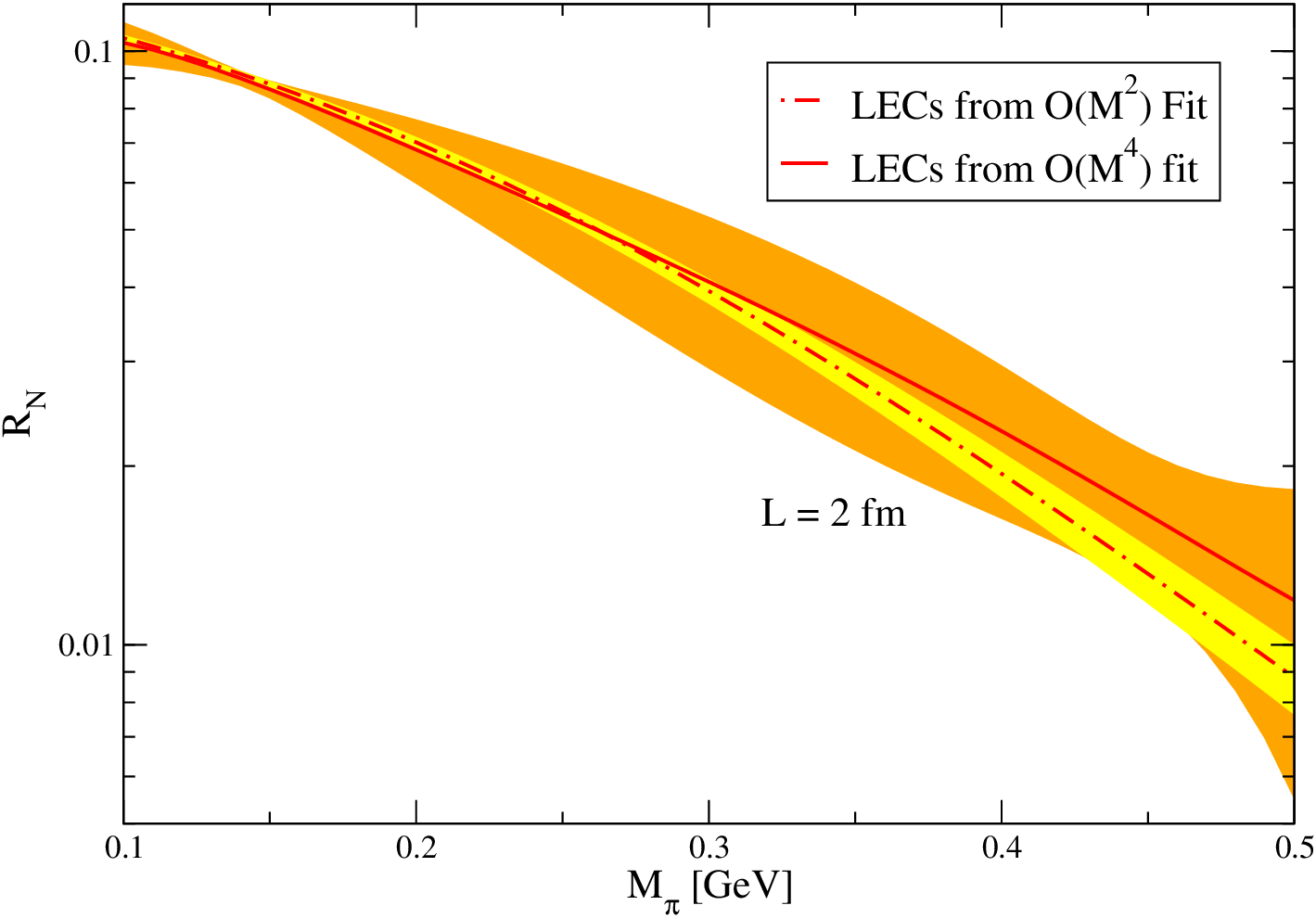}
\caption{\label{fig:RN}Relative finite volume effect for the nucleon mass as a
  function of the pion mass.}
\end{figure}

This situation makes the numerical study of finite volume effects for the
nucleon mass on the basis of Eq.~(\ref{eq:RNb}) problematic. We
proceed nonetheless and rely on a simple interpolation formula between the
physical and the lattice value (at around $M_\pi \sim 0.5$ GeV) of $g_A$ --
for all other quantities we rely on the chiral expansion. Our results are
shown in Fig.~\ref{fig:RN} and indicate that although with large
uncertainties, the finite volume corrections for the nucleon mass can be
calculated with this method. The figure contains two curves (with the
corresponding uncertainties): the dotdashed curve indicates what one
obtains if one stops the chiral expansion of the $\pi N$ scattering
amplitude at $O(p^2)$, whereas the solid line gives the result obtained
from the full $O(p^4)$ scattering amplitude. In the latter case in the
amplitude there are more LEC to be determined and correlations among them
are generated by the fits, but the net result is that the uncertainties
become larger than if one stops at $O(p^2)$. The comparison clearly shows
that the direct statistical estimate of the uncertainty at $O(p^2)$ is
quite optimistic.

\begin{figure}[t]
\includegraphics[width=0.48\textwidth]{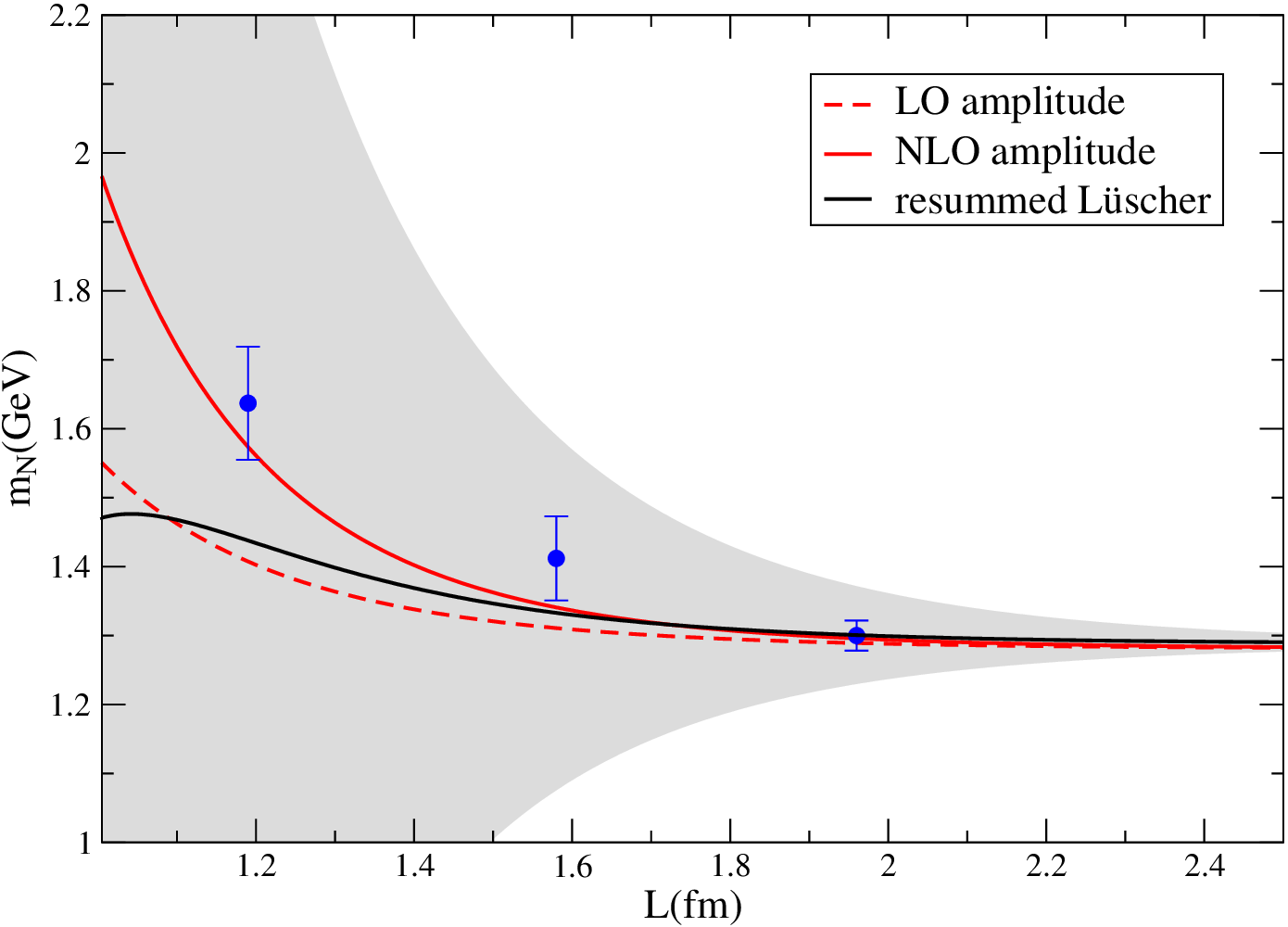}
\caption{\label{fig:MNL}Finite volume effect for the nucleon mass as a
  function of $L$ for $M_\pi=0.545$ GeV. The lattice data are from
  Ref.~\cite{Aoki:2002uc}. The shaded band represents the uncertainty of the
    calculation.}
\end{figure}

We conclude by comparing to the analysis in \cite{AliKhan:2003cu}. The
algebraic relation between the two approaches is fully understood: if we
expand the resummed L\"uscher formula Eq.~(\ref{eq:RNb}) to NLO we
reproduce exactly the result of Ref.~\cite{AliKhan:2003cu}. The partial
inclusion of higher orders obtained if we do not expand Eq.~(\ref{eq:RNb})
but insert the formulae of the $O(p^4)$ $\pi N$ scattering amplitude, on
the other hand, allows us to check the convergence of the series.
Fig.~\ref{fig:RN} shows a reasonable behaviour for $L=2$ fm and $M_\pi
\leq 0.5$ GeV. In Ref.~\cite{AliKhan:2003cu} the comparison between lattice
data and the CHPT calculation in finite volume has been performed even for
smaller volumes and higher pion masses and has shown a surprisingly good
agreement. The significance of this agreement can however only be assessed
if one has a good estimate of the uncertainties, something which had not
been attempted in~\cite{AliKhan:2003cu}. We can do this now and compare
our results to the same lattice data, which were obtained for $m_N$ at
$M_\pi \simeq 0.55$ GeV and volumes $L< 2$ fm -- a region where we would
not trust our formulae from the start. 
The comparison is shown in Fig.~\ref{fig:MNL} where one can see that if one
stops at NLO one goes indeed through the data, in agreement
with~\cite{AliKhan:2003cu}. The inclusion of higher orders, however, spoils
the good agreement, because for the central values of the LEC we have
obtained from our fit we come closer to the LO curve. The band which gives
our estimate of the uncertainty however shows that neither the agreement
nor the disagreement are of any significance, because the uncertainties at
these high quark masses and for such small volumes are simply too large.

\section{Conclusions}
We have discusssed the results of a full two-loop calculation of the pion
mass in finite volume in CHPT \cite{Haefeli:2005px}. We have shown that
contributions which are not included in the resummed asymptotic formula we
proposed some time ago \cite{Colangelo:2004sc,Colangelo:2005gd} are tiny
and can be neglected in the region of pion masses and volumes where CHPT
can be trusted. This check provides further evidence to our claim that the
resummed asymptotic formulae are the most efficient way to evaluate these
finite volume effects and in view of this there appears to be little need
to try and improve the analysis in \cite{Colangelo:2005gd} for masses and
decay constants of the pseudoscalar mesons.

The resummed formula can also be applied to non pseudo-Goldstone bosons,
and a particularly interesting application concerns the nucleon mass. This
has been calculated in finite volume to one loop in CHPT
\cite{AliKhan:2003cu} and compared to lattice data, although these had been
obtained for rather heavy pion masses and in small volumes. Somewhat
surprisingly, the CHPT calculation successfully described the volume
dependence of these data \cite{AliKhan:2003cu}. We have described here a
numerical analysis based on the resummed asymptotic formula and shown how
higher order corrections become more and more sizeable as the pion mass
increases. With the higher order corrections, also the uncertainties in the
calculation increase. Indeed we have shown that in the region of pion
masses and volumes where data are available the uncertainties are larger
than the effect itself and concluded that the observed agreement between
lattice data and the CHPT NLO calculation is accidental and of little
significance.

\section*{Acknowledgments}
G.C. thanks the organizers for the invitation to a very interesting and
perfectly organized workshop. We wish to thank Stephan D\"urr for useful
comments on the manuscript. This work has been supported by the
Schweizerischer Nationalfonds.


\begin{thebibliography}{39}
\bibitem{Gasser:1987zq}
  J.~Gasser and H.~Leutwyler,
  Nucl.\ Phys.\ B {\bf 307} (1988) 763.

\bibitem{Colangelo:2004sc}
  G.~Colangelo,
  Nucl.\ Phys.\ PS  {\bf 140} (2005) 120.

\bibitem{Bijnens:2005ne}
  J.~Bijnens, N.~Danielsson, K.~Ghorbani and T.~Lahde,
  arXiv:hep-lat/0509042.

\bibitem{Haefeli:2005px}
  C.~Haefeli,
  arXiv:hep-lat/0509078.\\
  G.~Colangelo and C.~Haefeli, in preparation.

\bibitem{Luscher:1985dn}
M.~L\"uscher,
Commun.\ Math.\ Phys.\ {\bf 104}, 177 (1986).

\bibitem{Colangelo:2004xr}
  G.~Colangelo and C.~Haefeli,
  Phys.\ Lett.\ B {\bf 590} (2004) 258.

\bibitem{Colangelo:2005gd}
  G.~Colangelo, S.~Durr and C.~Haefeli,
  Nucl.\ Phys.\ B {\bf 721} (2005) 136.

\bibitem{Colangelo:2002hy}
G.~Colangelo, S.~Durr and R.~Sommer,
  Nucl.\ Phys.\ Proc.\ Suppl.\  {\bf 119} (2003) 254.

\bibitem{Colangelo:2003hf}
  G.~Colangelo and S.~Durr,
  Eur.\ Phys.\ J.\ C {\bf 33} (2004) 543.

\bibitem{AliKhan:2003cu}
A.~Ali Khan {\it et al.} [QCDSF-UKQCD Collaboration],
Nucl.\ Phys.\ B {\bf 689}, 175 (2004) 

\bibitem{Hohler:1984ux}
  G.~Hohler and H.~Schopper, Springer (1983) (Landolt-Boernstein. New
  Series, I/9B2).  

\bibitem{Becher:2001hv}
  T.~Becher and H.~Leutwyler,
  JHEP {\bf 0106} (2001) 017.

\bibitem{Koma:2004wz}
  Y.~Koma and M.~Koma,
  Nucl.\ Phys.\ B {\bf 713} (2005) 575.

\bibitem{Aoki:2002uc}
  S.~Aoki {\it et al.}  [JLQCD Collaboration],
  Phys.\ Rev.\ D {\bf 68} (2003) 054502.

\bibitem{galatt}
  A.~A.~Khan {\it et al.},
  arXiv:hep-lat/0510061.\\
  R.~G.~Edwards {\it et al.}  [LHPC Collaboration],
  arXiv:hep-lat/0510062. The latter result has not
  yet been used in our fits.

\bibitem{CF}
G.~Colangelo and A.~Fuhrer, in preparation.

\bibitem{Fettes:1999wp}
  N.~Fettes, V.~Bernard and U.~G.~Meissner,
  Nucl.\ Phys.\ A {\bf 669} (2000) 269.

\bibitem{Hemmert:2003cb}
  T.~R.~Hemmert, M.~Procura and W.~Weise,
  Phys.\ Rev.\ D {\bf 68} (2003) 075009.

\bibitem{Kambor:1998pi}
  J.~Kambor and M.~Mojzis,
  JHEP {\bf 9904} (1999) 031.



\end{thebibliography}
\end{document}